\documentclass[PRL,aps,preprint,showpacs,amsmath,amssymb, superscriptaddress]{revtex4}

\usepackage{graphicx}
\usepackage{bm}% bold math
\usepackage{color}

\begin{document}

\title{Apparent first-order wetting and anomalous scaling in the two-dimensional Ising model}

\author{X.T. Wu}
\affiliation{Department of Physics, Beijing Normal University,
Beijing, 100875, China}
\affiliation{Institute for Theoretical Physics, KU Leuven, BE-3001 Leuven, Belgium}
\author{D. B. Abraham}
\affiliation{NYU-ECNU Institute of Mathematical Sciences at NYU Shanghai,
3663 Zhongshan Road North, Shanghai, 200062, China}
\author{J.O. Indekeu}
\affiliation{Institute for Theoretical Physics, KU Leuven, BE-3001 Leuven, Belgium}

\date{\today}% It is always \today, today,

\begin{abstract}
The global phase diagram of wetting in the two-dimensional ($2d$) Ising model is obtained through exact calculation of the surface excess free energy. Besides a surface field for inducing wetting, a surface-coupling enhancement is included. The wetting transition is critical (second order) for any finite ratio of surface coupling $J_{s}$ to bulk coupling $J $, and turns first order in the limit $J_s/J \rightarrow \infty$. However, for $J_s/J \gg 1$ the critical region is exponentially small and practically invisible to numerical studies. A distinct pre-asymptotic regime exists in which the transition displays first-order character. Surprisingly, in this regime the surface susceptibility and surface specific heat develop a divergence and show anomalous scaling with an exponent equal to 3/2. 
\end{abstract}


\maketitle

When a surface is exposed to an adsorbate at two-phase coexistence either droplets (partial wetting or ``nonwet") or a uniform layer (complete wetting or ``wet") of one of the phases may form on it. Delicate tuning of surface or bulk properties may allow one to achieve a surface phase transition or critical phenomenon from partial to complete wetting. The wetting transition, so called, has been studied experimentally and theoretically for some 35 years now; for reviews, see, e.g.,  \cite{Gen,Fish,Sul,Diet,Bonn}. The first exact solution beyond mean-field theory revealed a critical wetting transition (of second order) in the 2$d$-Ising model with a surface field \cite{A1,A2}. When antiferromagnetic surface couplings are added, critical wetting persists \cite{Abraham1988}. However, for strong ferromagnetic surface couplings new physics arises, as we show in this Letter. 

Monte Carlo simulations of wetting in the 3$d$-Ising model with a surface field {\em and} a surface-coupling enhancement have unveiled a rich global phase diagram, featuring first-order and critical wetting, separated by tricritical wetting \cite{Binder}, in accord with qualitative predictions from Landau theory \cite{Nak}. In $d=2$, however, where thermal fluctuation effects on wetting are pronounced, {\em only critical wetting transitions}, belonging to a single universality class, are expected \cite{KLZ1,KLZ2,Fish}. Nevertheless, an exact calculation revealed that first-order wetting is possible when an extra defect line is introduced \cite{Forgacs}. Furthermore, numerical evidence for first-order wetting was found in Monte Carlo simulations of the 2$d$ Ising model with an extra spin state (a vacancy) \cite{Cotes,Trobo, Albano}.  

We investigate the global phase diagram for wetting in $d=2$ for short-range forces and answer the following fundamental questions. Is first-order wetting possible in $d=2$ for the standard spin-1/2 Ising model with a surface, by enhancing the spin-spin coupling at the surface? What is the precise character of the wetting transition in $d=2$; in particular, how wide is the critical region and are there distinct pre-asymptotic regimes? 

Consider a set of Ising spins $\sigma (n,m)=\pm 1$ located at
points $(n, m)$ of the planar square lattice $\Lambda(n,m)$ such
that $1 \leq n \leq N$, $ 1\leq m\leq M$. The energy of a configuration
$\{\sigma \}$ of spins is given by
\begin{eqnarray}
E(\{ \sigma \}) & = & -\sum_{m=1}^{M} \{  H_1(m) \,\sigma(1,m)+H_N(m) \,\sigma(N,m) \} \nonumber \\& & -\sum_{m=1}^{M}  J_0 \,\sigma(1,m)\sigma(2,m) -\sum_{m=1}^{M} J_s\, \sigma(2,m)\sigma(2,m+1) \nonumber \\& & -\sum_{m=1}^{M} \sum_{n=2}^{N-1}  J_1\, \sigma(n,m)\sigma(n+1,m) -\sum_{m=1}^{M} \sum_{n=3}^{N-1} J_2 \, \sigma(n,m)\sigma(n,m+1)
\label{hamiltonian}
\end{eqnarray}
The fields $H_1(m)$ and $H_N(m)$ allow us to fix boundary spins.  The  spin-spin coupling $J_0$ ($> 0$) acts as an {\em effective surface field} on the first layer ($n=2$) of  free spins; this is the usual ``wetting" term, which for mixtures corresponds to a differential surface fugacity. We denote by $h_1 \equiv \beta J_0$ the absolute value of the (reduced) surface field, where $\beta=1/k_BT$, with $k_B$ the Boltzmann constant
and $T$ the absolute temperature. Modified spin-spin couplings $J_s$ along the surface take into account the changed environment of molecular interactions at the surface in such a way that exact solution is still possible. $J_1$ and $J_2$ are the usual (ferromagnetic) ``bulk" nearest-neighbour  couplings.  Periodic boundary conditions $\sigma(n,M+1) = \sigma(n,1 )$, which we impose, are essential to generate exact solutions. The normalized canonical probability is $ P(\{ \sigma \})=Z^{-1}\exp [-\beta E]$,
where $Z$ is the partition function. 

We will make use of two types of ``wall" boundary conditions:
\begin{equation}
A:H_1(m), H_N(m)= +\infty, \; \mbox{for} \;1 \leq m \leq M
\label{boundarya}
\end{equation}
which, for $T\leq T_c$, with $T_c$ the bulk critical temperature, force a state with positive spontaneous bulk magnetization, in the thermodynamic limit $M\rightarrow \infty$ followed
by $N \rightarrow \infty$; and
\begin{eqnarray}
B: H_1(m) &=&
\begin{cases} -\infty,  \; \mbox{for} \; 1 \leq m \leq S\; \\ +\infty, \; \mbox{for} \; S<m\leq M
\end{cases}\nonumber \\
 H_N(m)& = &  +\infty, \; \mbox{for} \; 1 \leq m \leq M, 
\label{boundaryb}
\end{eqnarray}
which force a long contour of surface length $S$, beginning at $(1, \frac{1}{2})$ and ending
at $(1,S+\frac{1}{2})$ which delimits the region of predominantly
negative magnetization.

The surface excess free energy $f$ (per unit length of surface, $S$) can be obtained from
\begin{equation}\label{free}
\beta f=-\lim_{S \rightarrow \infty}\lim_{\Lambda \rightarrow \infty}\frac{1}{S} \ln \frac{Z_{B}}{Z_{A}}
\end{equation}
where partition functions $Z_{A}$ and $Z_{B}$ correspond to the respective boundary conditions. In the language of wetting phenomena this definition ensures that $f$ equals $\gamma_{+-}\cos \theta_Y$ in the nonwet state and $\gamma_{+-}$ in the wet state ($\theta_Y=0$), where $\theta_Y$ is Young's contact angle and $\gamma_{+-}$ is the surface tension of a free interface between $+$ and $-$ phases in bulk. We obtain the analytic form
\begin{equation}\label{Zratio}
\frac{Z_B}{Z_A}=\frac{i}{2\pi } \int_0^{2\pi }d\omega \;e^{iS\omega}\tan \delta^*(\omega /2)\frac{(e^{\gamma(\omega)}-Q_+)(e^{\gamma(\omega)}-Q_-)}{(e^{\gamma(\omega)}-P_+)(e^{\gamma(\omega)}-P_-)}
\end{equation}
where $\gamma (\omega)$, $\delta^* (\omega)$ are elements of the Onsager hyperbolic triangle:
\begin{equation}
\cosh \gamma (\omega)=\cosh 2K_1^* \cosh 2K_2 -2\sinh 2K_1^* \sinh K_2 \cos \omega
\end{equation}
and
\begin{equation}
\cosh 2K_1^*=\cosh 2K_2 \cosh \gamma (\omega)-\sinh K_2 \sinh \gamma (\omega) \cos \delta^* (\omega),
\end{equation}
with $K_i \equiv \beta J_i$, and with dual couplings $K_i^*$ satisfying $\tanh K_i^*=e^{-2K_i}$, for $i=1,2$.
The quantities $P_{\pm}$ and $Q_{\pm}$ are real-valued and independent of $\omega$.

The integrand is singular at values of $\omega$ for which $
e^{\gamma(\omega)}=P_{\pm}$.
The results for $P_{\pm}$ are
\begin{equation}\label{P}
P_{\pm}=\frac{s\pm \sqrt{s^2-r^2+1}}{r+1},
\end{equation}
where, defining $2 K'_2 \equiv 4 K_s-2 K_2$ with $K_s \equiv \beta J_s$,
\begin{equation} \label{r}
r=\frac{e^{2K'_2}-\cosh 2K_2}{\sinh 2K_2}
\end{equation}
and
\begin{equation}
s=\cosh 2K_1^* \frac{e^{2K'_2}\cosh 2K_2-1}{\sinh 2K_2}-e^{2K'_2}\sinh 2K_1^* \cosh 2h_1
\end{equation}
The $Q_{\pm}$ have similar structure to the $P_{\pm}$ but, crucially, never coincide with the $P_{\pm}$. Hence they cannot remove the simple poles coming from the zeros in the denominator of \eqref{Zratio}, needed to establish the limiting free energy. The details of $Q_{\pm}$ do not contribute to the location of the poles but only to the residues, and will be given elsewhere.

Henceforth we assume isotropy in bulk, $J_1=J_2=J$, so $K_1=K_2=K$. The singularity is given by $P_+=1$,
and for $P_+>1$ (nonwet state), $ f$ is obtained through
\begin{eqnarray}\label{coshbetaf}
\cosh \beta f  =  \cosh(2K-2K^*)+1-\frac{1}{2}(P_++\frac{1}{P_+}),
\end{eqnarray}
while for $P_+<1$ or complex $P_+$ (wet state), $\beta f= 2K+\ln \tanh K$, which equals $\beta \gamma_{+-}$ \cite{O,RW}. 

For a given $K$ we denote the value of $h_1$ at wetting by $h_{1w} $. For the special case $J_s=J$ solved in 1980 the critical wetting phase boundary satisfies $e^{2K}(\cosh 2K-\cosh 2h_{1w})=\sinh2K$ \cite{A1,A2}.
\begin{figure}
 \begin{center}
    \resizebox{10cm}{8cm}{\includegraphics{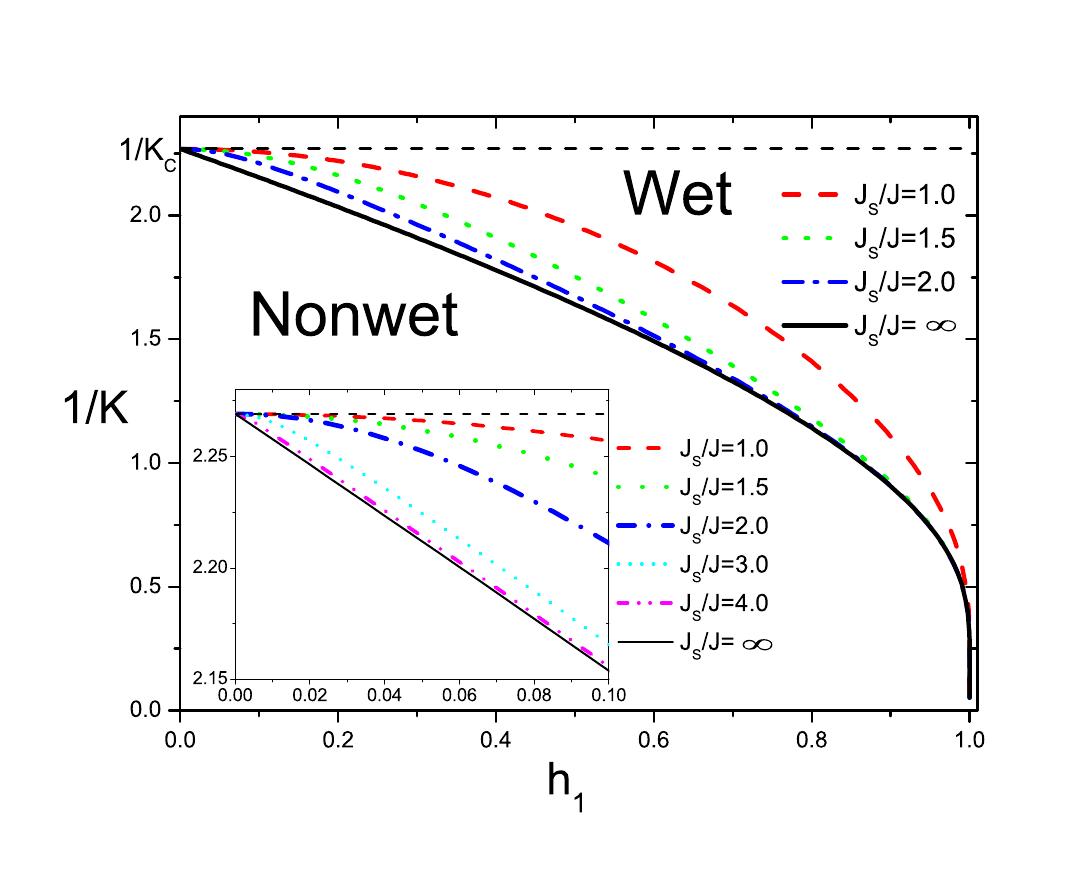}}
  \end{center}
\vskip -1cm

\caption{Wetting phase boundaries in surface field $h_1$ and temperature $1/K$, for various surface-coupling enhancements $J_s/J$. For finite $J_s/J$ the wetting transition is of second order and the phase boundary is parabolic near $h_1=0$, whereas for $J_s/J = \infty$ the wetting transition is of first order and the phase boundary is linear near $h_1=0$. The horizontal dashed line marks bulk criticality.}
\end{figure}
Fig.1 shows critical wetting phase boundaries for $J_s \geq J$. For $J_s/J=\infty$ we obtain $
\cosh ^2 2K/\sinh 2K-\cosh 2h_{1w}=1$, which simplifies to
\begin{equation}\label{cwpblimit}
h_{1w}=K-K^*= K+\frac{1}{2}\ln \tanh K,\; \;\;\mbox{for} \; J_s/J=\infty,
\end{equation}
and has a simple physical interpretation. For $J_s/J \gg 1$ (surface ferromagnetic limit) the surface magnetization ${\hat m}_1$ (at $n=2$ on the lattice) saturates to +1 or -1, since all surface spins are aligned. The wetting transition is induced by a massive surface spin flip from -1 to +1, causing an interface between + and - phases in bulk to unbind from the surface. Anticipating a {\em first-order} transition for $J_s/J \rightarrow \infty$, we can conjecture $h_{1w}$ simply by equating the surface energy gain of wetting to the surface tension cost of a free interface. 

The phase boundary for $J_s/J=\infty$ is linear near the bulk critical point. For $K \rightarrow K_c$, $h_{1w} \sim 2 (1-K_c/K)$,
where $K_c=\frac{1}{2}\ln (1+\sqrt{2}) \approx 0.4407$ is the bulk critical coupling. This differs from the quadratic (or higher-order) behaviour found for the critical wetting phase boundary near bulk $T_c$ for finite $J_s/J$. The linear character is reminiscent of mean-field first-order wetting near surface-bulk multicriticality, with tricritical wetting for $T \rightarrow T_c$ \cite{Nak}. 

\begin{figure}
 \begin{center}
    \resizebox{13cm}{11cm}{\includegraphics{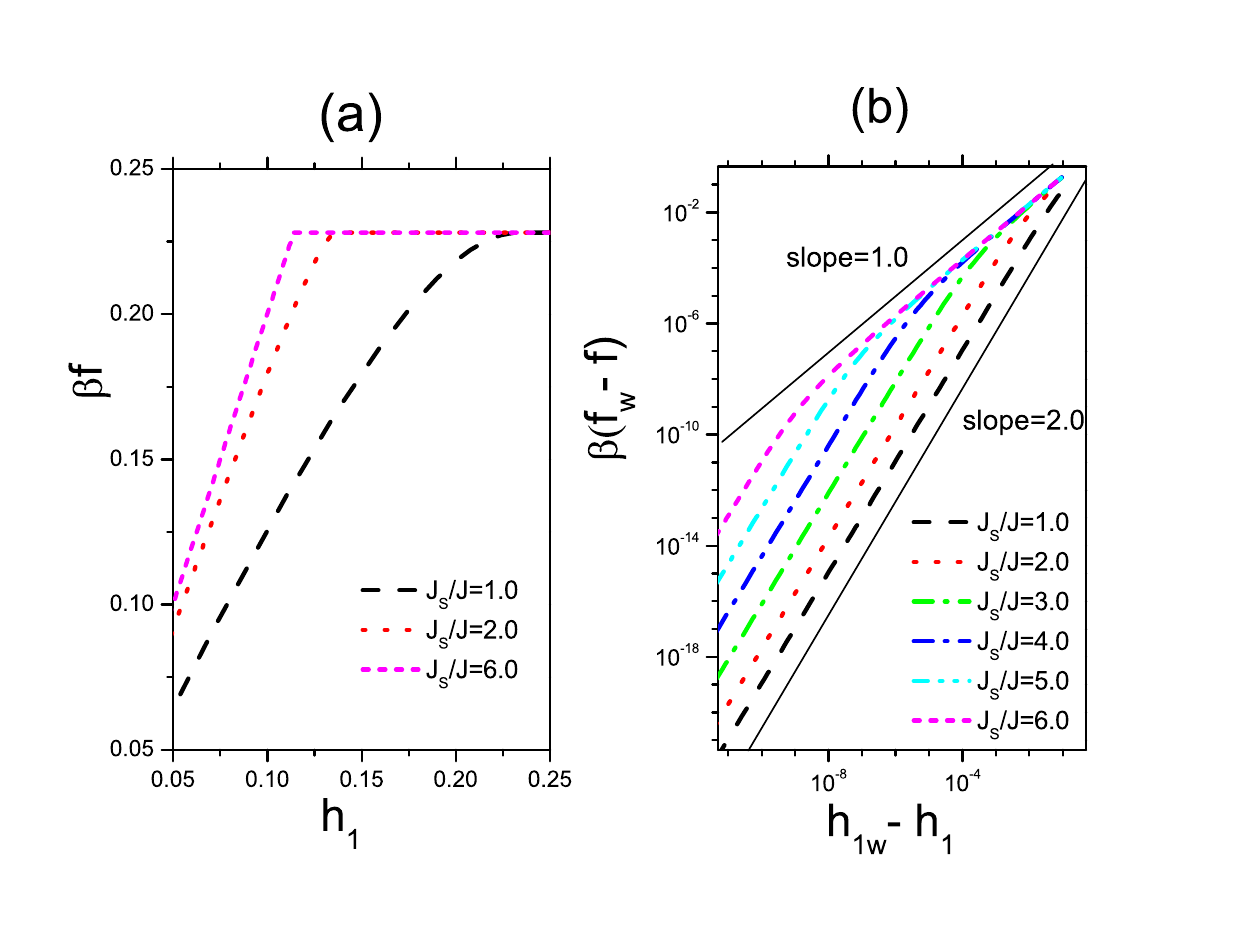}}
  \end{center}
\vskip -1cm
  
\caption{Surface excess free energy $ f$ versus surface field $h_1$. (a) The transformation of  a parabolic singularity (second-order transition)  into a corner (apparent first-order transition) as $J_s/J$ is increased. (b) The crossover from apparent first-order to asymptotic second-order character in the free-energy singularity, in a log-log plot. Solid lines with slopes 1 and 2 have been added (thin, black). The temperature $T  (< T_c)$
is fixed through $1/K = 2$. }
\end{figure}

Remarkably, the wetting transition already appears first order at large but finite $J_s/J$. Fig.2 shows the surface excess free energy $f$ near the transition.  We fix the temperature through $1/K=2$ and vary  $h_1$. In Fig.2a, there clearly appears a sharp corner for $J_s/J =6$, suggesting first-order behavior. The singular part of $f$ is shown in Fig.2b. We denote the value of $f$ at wetting by $f_w \equiv f(h_{1w})$. The simple behavior $f_w-f\propto (h_{1w}-h_1)^2$ found for $J_s/J=1$ is the signature of second-order wetting. However, for larger surface coupling, say $J_s/J > 4$, there is an extended range of $h_1$ for which $f_w-f\propto (h_{1w}-h_1)$, indicating first-order character. Only very near the transition does $f_w-f$ cross over from $(h_{1w}-h_1)$ to $ (h_{1w}-h_1)^2$-like behavior. In the regime where $f_w-f\propto (h_{1w}-h_1)$, the transition is effectively of first order.

\begin{figure}
 \begin{center}
    \resizebox{13cm}{11cm}{\includegraphics{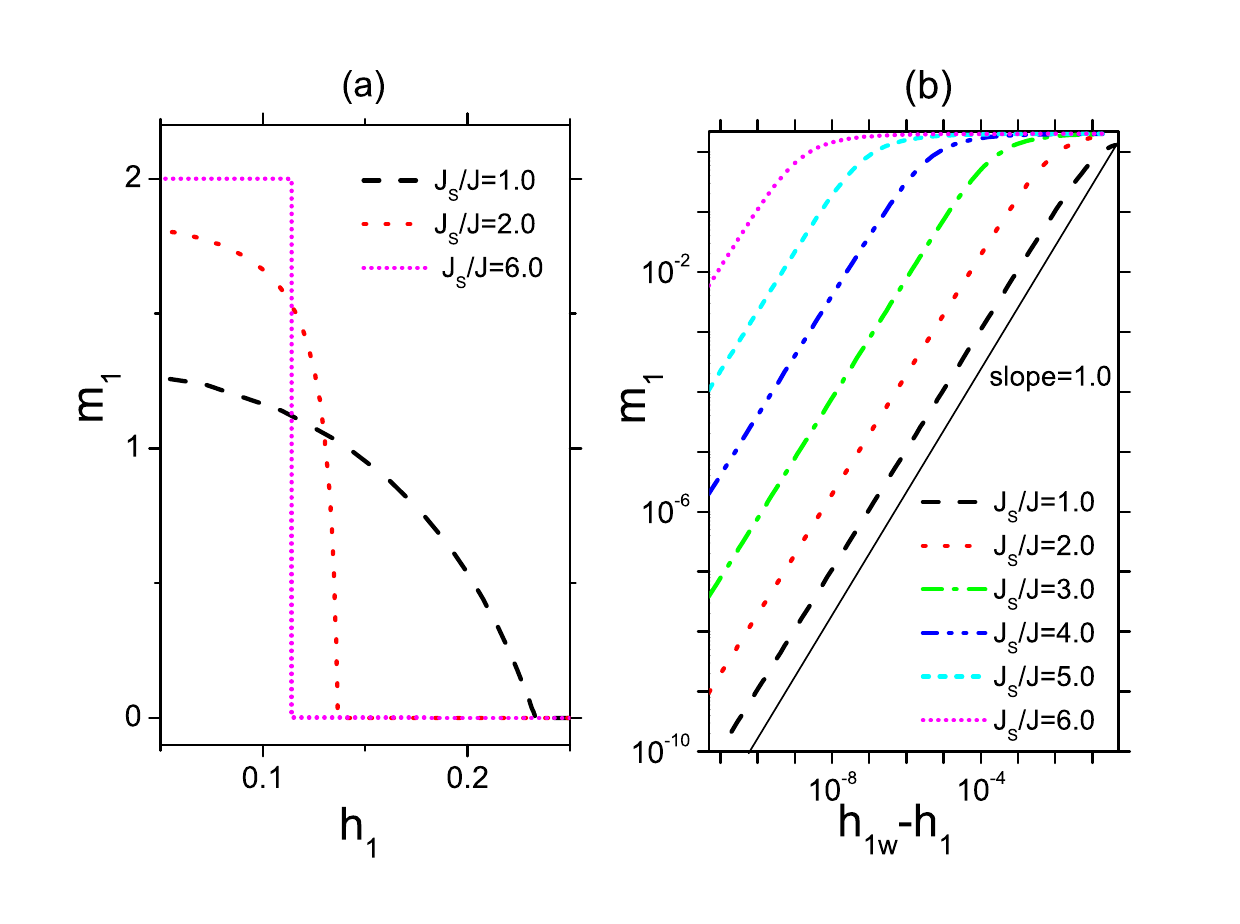}}
  \end{center}
\vskip -1cm
  
\caption{Surface excess magnetization $ m_1$ versus surface field $h_1$. (a) The linear approach to zero for (second-order) critical wetting gradually transforms into an apparent discontinuity as $J_s/J$ is increased, indicating first-order character. (b) The crossover from first-order (piecewise constant) to second-order character (vanishing with critical exponent 1), in $m_1$ versus surface field in a log-log plot. A solid line with slope 1 has been added (thin, black).}
\end{figure}

\begin{figure}
 \begin{center}
    \resizebox{13cm}{11cm}{\includegraphics{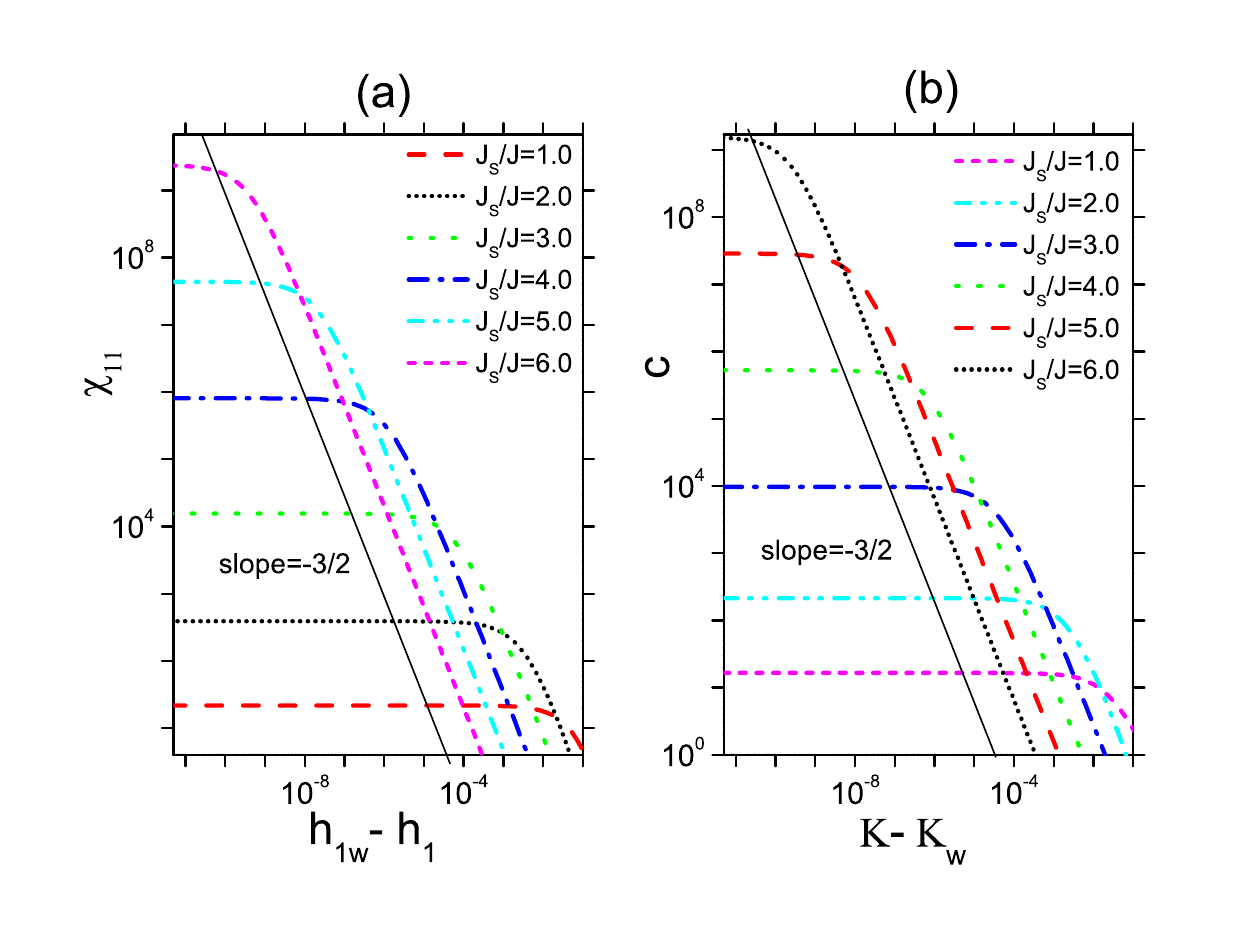}}
  \end{center}
\vskip -1cm
  
\caption{Anomalous scaling of (a) the  surface susceptibility $\chi _{11}$ and (b) the surface specific heat $c$.  The apparent divergence, with exponent $3/2$, is manifest and persists until the critical region is reached. There, a crossover to a constant value takes effect. This value is the magnitude of the jump in the thermodynamic response function at critical wetting. Note how the critical region shrinks as $J_s/J$ is increased (cf.Eq.\eqref{critregion}). Solid lines with slope -3/2 have been added (thin, black).}
\end{figure}

The emerging first-order character is conspicuous in the surface {\em excess} magnetization $m_1$, defined as 
$m_1\equiv -\beta (\partial f/\partial h_1)$ and related to the surface magnetization ${\hat m}_1$ through $m_1 = {\hat m}_1 +1$, so that $m_1=0$ in the wet state.
Fig.3 shows $m_1$ for $1/K=2$. In Fig.3a,
$m_1$ develops a step-like singularity as $J_s/J$ is increased. Fig.3b shows detail near the transition point. For large $J_s/J$, $m_1$ stays constant (at $m_1 = 2$) until very close to the transition, and eventually crosses over to the second-order transition behavior, which is a linear decrease $m_1 \propto (h_{1w}-h_1)$, corresponding to lines of slope $1$  in Fig.3b.

New physics arises when examining the surface susceptibility and the surface specific heat. Accompanying the emerging first-order character, there is anomalous scaling  in the surface susceptibility $\chi _{11}$,
 defined as the second derivative of $f$ with respect to $h_1$.
Fig.4a shows $\chi _{11}$ for different $J_s/J$ at $1/K=2$. For the standard second-order wetting transition $\chi _{11}$
makes a finite jump. For large $J_s/J$ the jump is still finite but very large. Near the transition point $\chi _{11}$  displays an apparent {\em divergence }
according to a power law as $h_1$ approaches $h_{1w}$ from below (nonwet state). For example, for $J_s/J=6$ we find $\chi _{11} \propto (h_{1w}-h_1)^{-3/2}$ for $10^{-9} <h_{1w}-h_1<10^{-3}$, implying an effective exponent for $\chi _{11}$ equal to $3/2$. This cannot be explained by the usual scaling relations. A similar anomaly is found for the surface specific heat $c$, which is proportional to the second derivative of $f$ with respect to $1/K$. Fig.4b shows $c$ for different $J_s/J$, with $h_1$ fixed at the value of $h_{1w}$ found for $1/K=2$. The exponent characterizing the apparent divergence of $c$ at wetting, for large $J_s/J$, also equals $3/2$.

We now demonstrate the robustness of the linear dependence of $f$ on $h_1$ near the wetting transition for $J_s/J\gg 1$ and explain the anomalous scaling. For $J_s/J\gg 1$, we have $K'_2 \sim 2 (J_s/J) K$, so $r \gg 1$ in view of \eqref{r}.
We fix $K$ and vary $h_1$. At the transition, $
r=s(h_{1w})$.
We expand $s$ about $h_{1w}$, with $1 \gg \Delta h_1 \equiv h_{1w}-h_1 > 0$,
\begin{eqnarray}
s(h_{1w}-\Delta h_1) =  r \left\{1+2\Delta h_1\sinh2h_{1w} + {\cal O}((\Delta h_1)^2)+ {\cal O}(\Delta h_1/r)\right \}
\end{eqnarray}
The form \eqref{P} of $P_+$ suggests two important scaling limits. The first is the {\em critical} limit $r^2 (s/r-1) \ll 1$, to which we return later. The second is the {\em strong surface coupling} limit $r^2 (s/r-1) \gg 1$, or, $1\gg \Delta h_1 \gg e^{-8(J_s/J)K} = e^{-8K_s}$. In this limit we get
\begin{equation}
P_+ =  1+\sqrt{2\Delta h_1\sinh2h_{1w}}-1/r + {\cal O}((\Delta h_1)^2)+ {\cal O}(\Delta h_1/r) +{\cal O}( 1/r^2).
\end{equation}

The free energy difference $f_w - f$ in this limit is interesting. 
Using \eqref{coshbetaf}
we obtain
the surprising form
\begin{equation}\label{unusualf}
\beta (f_w-f) = 2 \Delta h_1-\frac{2}{r}\sqrt{\frac{\Delta h_1}{\sinh 2h_{1w}}}+ {\cal O}((\Delta h_1)^2)+ {\cal O}(\Delta h_1/r) +{\cal O}( 1/r^2)
\end{equation}
implying that the transition is {\em effectively of first order}, since the second term is much smaller than the first due to the prefactor $1/r$. However, this nonlinear correction term becomes all-important when taking the second derivative of the free energy! Thus, \eqref{unusualf} allows one to capture instantly the anomalous scaling for the surface susceptibility,
\begin{equation}
\chi_{11} \propto (\Delta h_1)^{-3/2}
\end{equation}

In the ``temperature" direction, we can get similar but more complicated expansions. If we fix $h_1$ and $J_s/J$, and expand the free energy about the wetting point $K=K_w$, we obtain, with $\Delta K \equiv K-K_w > 0$,
\begin{equation}
\beta (f_w-f) \approx \frac{A}{2\sinh 2 h_{1}}\Delta K-\frac{1}{2r\sinh 2h_{1}}\sqrt{2A\Delta K},
\end{equation}
where $
A \equiv  {\cal A}(K_w,h_{1}, J_s/J) \equiv  \partial \ln (s/r)/\partial K |_{K=K_w}$. This clarifies why there is anomalous scaling also in the specific heat, with the same exponent $3/2$, as illustrated in Fig.4b.

No matter how large $J_s/J$, the asymptotic behavior in the limit $\Delta h_1 \rightarrow 0$ is invariably critical wetting (second order transition). The only exception is $J_s/J = \infty$, for which  \eqref{unusualf} holds exactly with $1/r = 0$.  The asymptotic behavior for all finite $J_s/J$ is easily obtained in the critical scaling limit $r^2 (s/r-1) \ll 1$, with the result $f_w-f \propto (\Delta h_{1} )^2$. However, for large $J_s/J$ the critical region is {\em exponentially small}, i.e.,
\begin{equation}\label{critregion}
0 \leq \Delta h_{1}  \ll e^{-8 (J_s/J)K}.
\end{equation}

The global wetting phase diagram at bulk coexistence, in the variables $h_1$ and $J_s/J$, featuring the surface ferromagnetic as well as antiferromagnetic regime, is presented in Fig.5 for a representative fixed temperature $1/K = 2$. The  phase boundary separating the wet and non-wet regions (thick solid line) consists of critical wetting  but develops apparent first-order character for large $J_s/J$. True first-order wetting is obtained for $J_s/J = \infty$.

\begin{figure}
 \begin{center}
    \resizebox{11cm}{11cm}{\includegraphics{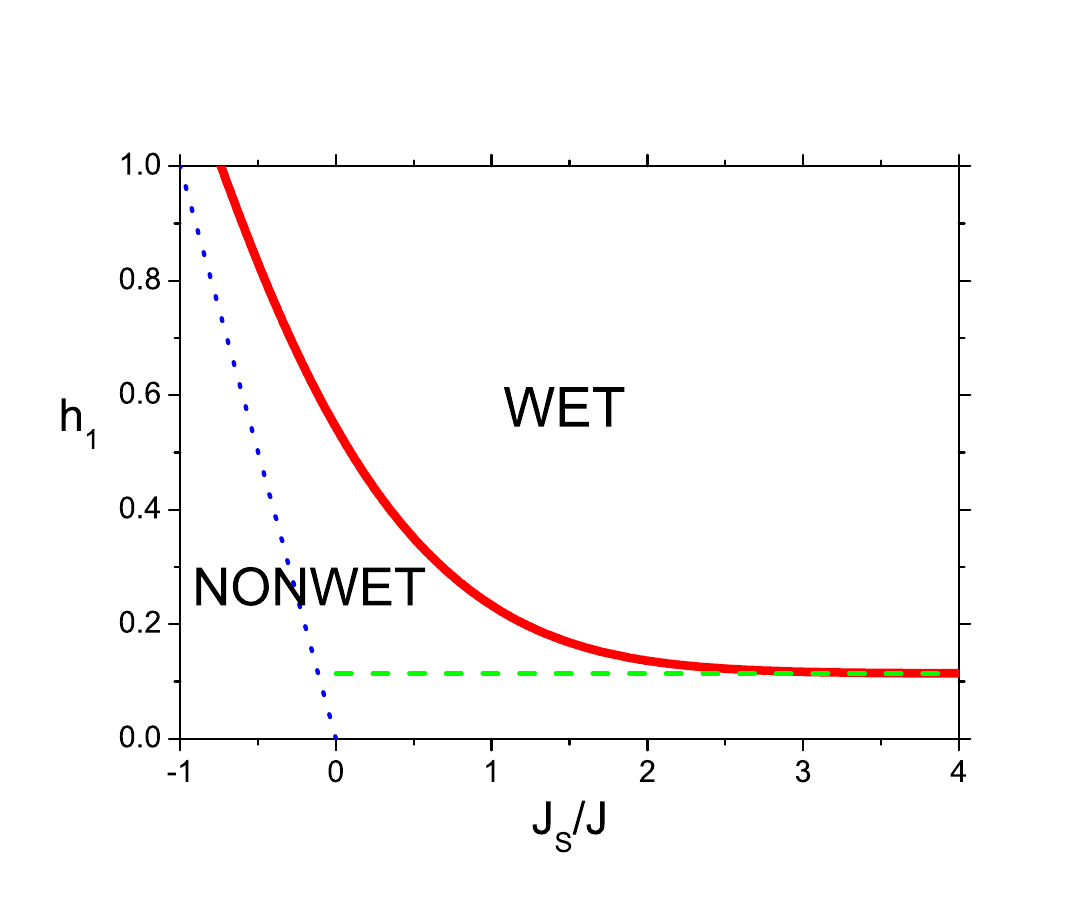}}
  \end{center}
\vskip -1cm
  
\caption{Global wetting phase diagram at bulk coexistence in surface field $h_1$ and surface-coupling enhancement $J_s/J$, for $1/K=2$. Wet and nonwet regions are separated by critical wetting (thick solid line; red), which develops first-order character when approaching the would-be first-order wetting line $h_{1w} = 0.11403...$ derived in the large $J_s/J$ limit (dashed line; green). The thin dotted line (blue) is parallel to the asymptote to the critical wetting line in the strongly antiferromagnetic surface regime. }
\end{figure}

For $J_s<0$ (antiferromagnetic surface coupling) and large $|J_s/J|$ the surface forms a perfect antiferromagnetic chain, and ${\hat m}_1=0$, unless the (uniform) surface field is strong enough to break the staggered surface order. Depinning becomes possible for $K_0 > -2 J_s$, or $h_1 > -2 (J_s/J) K$.
This defines the {\em slope} of the asymptote for large $|J_s/J|$ to the critical wetting phase boundary found for $J_s < 0$. 

The apparent first-order character of the wetting transition for large $J_s/J$ can be interpreted physically. In the Solid-on-Solid  model description of interface delocalization in $d=2$, the interface unbinds from the surface in a continuous and gradual manner for $J_s/J \lesssim 1$, while for $J_s/J \gg 1$ it can unbind only via quantum tunneling through a high activation barrier \cite{KLZ1,Fish}. This can explain an {\em effective} first-order wetting transition, which crosses over to a continuous one only extremely close to the transition. 

At large $J_s/J$ the ultimate crossover to critical wetting cannot be detected by accurate numerical techniques for finite systems (e.g., such as those developed in \cite{wu1,XWJSP}) and it is unrealistic to expect that it could be seen in Monte Carlo simulation, or in an experiment, in a (quasi-)$2d$ system. The effective first-order transition with novel scaling properties, which we have highlighted, is for all practical purposes the dominant wetting behavior.

In conclusion, by exact solution we have shown that the wetting transition in the 2$d$-Ising model is critical for all finite $J_s/J$, but displays first-order character for large $J_s/J$.  This apparent first-order behavior is accompanied by anomalous scaling of the surface susceptibility and the surface specific heat, featuring for both quantities an apparent divergence with an exponent 3/2. 

We acknowledge KU Leuven Research Grant OT/11/063 and thank A.O. Parry, D. Huse, M. Kardar and K. Binder for comments on our manuscript. X.T. Wu is supported by the National Science Foundation of China (NSFC) under Grant No. 1175018. D.B. A. acknowledges the support of the NYU-ECNU Institute of Mathematical Sciences at NYU Shanghai. J.O. I. thanks Mehran Kardar for hospitality at MIT, Cambridge, Massachusetts.


\begin{thebibliography}{}
\bibitem{Gen} P.-G. de Gennes, Rev. Mod. Phys. {\bf 57}, 827 (1985)
\bibitem{Fish} M.E. Fisher, J. Chem. Soc., Faraday Trans. 2, {\bf 82}, 1569 (1986)
\bibitem{Sul} D.E. Sullivan and M.M. Telo da Gama, in {\em Fluid interfacial phenomena}, ed. C.A. Croxton (Wiley, New York, 1986), p.45
\bibitem{Diet} S. Dietrich, in {\em Phase transitions and critical phenomena}, ed. C. Domb and J.L. Lebowitz (Academic, London, 1988), vol.12, p.1
\bibitem{Bonn} D. Bonn, J. Eggers, J.O. Indekeu, J. Meunier and E. Rolley, Rev. Mod. Phys. {\bf 81}, 739 (2009)
\bibitem{A1} D.B. Abraham, Phys. Rev. Lett., {\textbf 44}, 1165  (1980)
\bibitem{A2} D.B. Abraham, in {\em Phase transitions and critical phenomena}, ed. C. Domb and J.L. Lebowitz (Academic, London, 1986), vol.10, p.1.
\bibitem{Abraham1988} D. B. Abraham, L. F. Ko, and N. M. \u{S}vraki\'c, Phys. Rev. B (R) {\bf 37}, 9871 (1988).
\bibitem{Binder} K. Binder and D.P. Landau, Phys. Rev. B {\bf 37}, 1745 (1988); E.V. Albano and K. Binder, Phys. Rev. Lett. {\bf 109}, 036101 (2012)
\bibitem{Nak} H. Nakanishi and M.E. Fisher, Phys. Rev. Lett. {\bf 49}, 1565 (1982)
\bibitem{KLZ1} R. Lipowsky, D. M. Kroll and R.K.P. Zia, Phys. Rev. B {\bf 27}, 4499 (1983).
\bibitem{KLZ2} D.M. Kroll, R. Lipowsky and R.K.P. Zia, Phys. Rev. B(R) {\bf 32} 1862 (1985).

\bibitem{Forgacs} G. Forgacs, N.M. \u{S}vraki\'c and V. Privman, Phys. Rev. B {\bf 37}, 3818 and {\bf 38}, 8996 (1988).
\bibitem{Cotes} S.M. Cotes and E.V. Albano, Phys. Rev. E {\bf 83}, 061105 (2011).
\bibitem{Trobo} M.L. Trobo, E.V. Albano and K. Binder, Phys. Rev. E {\bf 90}, 022406 (2014).
\bibitem{Albano} E.V. Albano and K. Binder, J. Stat. Phys. {\bf 157}, 436 (2014).
\bibitem{O} L. Onsager, Phys. Rev. {\bf 65}, 117 (1944).
\bibitem{RW} C. Rottman and M. Wortis, Phys. Rev. B {\bf 24}, 6274 (1981).


\bibitem{wu1} X.T. Wu and J. Y. Zhao, Phys. Rev. B {\bf 80}, 104402
(2009).
\bibitem{XWJSP} X.T. Wu, J. Stat. Phys. {\bf 157}, 1284 (2014).

\end{thebibliography}
\end{document}